\documentstyle[twocolumn,aps]{revtex}
\begin{document}
\input psfig
\draft

%
\twocolumn[\hsize\textwidth\columnwidth\hsize\csname@twocolumnfalse\endcsname
%
%

\title{Robust D-wave Pairing Correlations
\\ in a Hole-Doped 
Spin-Fermion Model for Cuprates}

\author{Mohammad Moraghebi$^1$, Seiji Yunoki$^2$ and Adriana Moreo$^1$}

\address{$^1$Department of Physics, National High Magnetic Field Lab and
MARTECH,\\ Florida State University, Tallahassee, FL 32306, USA}

\address{$^2$Materials Science Center, University of Groningen,
Nijenborgh 4, 9747 AG Groningen, The Netherlands}

\date{\today}
\maketitle

\begin{abstract}

Pairing correlations are studied numerically in the hole-doped 
spin-fermion model for cuprates. Simulations performed on up to 
$12 \times 12$ clusters provide robust 
indications of 
D-wave superconductivity
away from half-filling. The pairing correlations are the strongest in the 
direction {\it perpendicular} to the dynamic stripe-like inhomogeneities 
that appear in the ground state at some densities. 
An optimal doping, where the correlations reach a
maximum value, was observed at about 25\% doping, 
in qualitative agreement with high $T_c$ cuprates' experiments. 
On the other hand, pairing correlations are suppressed by
static stripe inhomogeneities.

\end{abstract}

\pacs{PACS numbers: 74.20.De, 74.20.Rp, 74.72.-h}
\vskip2pc]
\narrowtext

The nature of high temperature superconductors is an important open
problem in the area of strongly correlated electrons \cite{Kar}. 
In this context,
models for cuprates have been extensively used to search for
superconductivity (SC) arising from a purely electronic
mechanism. In spite of this effort, 
the current situation is still confusing, with few
positive reports of ground-state SC in electronic models 
using truly unbiased many-body techniques. 
Recently, a simple spin-fermion (SF) model has been 
proposed for cuprates \cite{adri1}.
The main advantage of the SF-model is its simplicity
for numerical studies, while still keeping a realistic, unbiased, 
and non-trivial character. Previous
studies have already shown that several of its properties, 
such as magnetic incommensurability, existence of a density-of-states
pseudogap at the chemical potential, and the 
shape of the Fermi surface, resemble experimental data for high $T_c$
cuprates \cite{adri1}. In addition, upon hole doping, the 
ground state 
exhibits charge incommensurability due to the formation of
hole-rich vertical and horizontal stripes \cite{adri1}. 
The goal of this paper is to study for the first time the
pairing correlations of this model, and determine the role that 
charge stripes play in the pairing process. To our surprise, relatively robust
pairing correlations in the D-wave channel were detected, correlated
with the presence of dynamical stripes.

The SF-model is constructed as an interacting system of
electrons and spins, mimicking phenomenologically the
coexistence of charge and spin degrees of freedom in 
the cuprates \cite{Pines}. Its Hamiltonian is given by
$$
{\rm H=
-t{ \sum_{\langle {\bf ij} \rangle\alpha}(c^{\dagger}_{{\bf i}\alpha}
c_{{\bf j}\alpha}+h.c.)}}
+{\rm J
\sum_{{\bf i}}
{\bf{s}}_{\bf i}\cdot{\bf{S}}_{\bf i}
+J'\sum_{\langle {\bf ij} \rangle}{\bf{S}}_{\bf i} \cdot{\bf{S}}_{\bf j}},
\eqno(1)
$$
\noindent where ${\rm c^{\dagger}_{{\bf i}\alpha} }$ creates an electron
at site ${\bf i}=({\rm i_x,i_y})$ with spin projection $\alpha$,  
${\bf s_i}$=$\rm \sum_{\alpha\beta} 
c^{\dagger}_{{\bf i}\alpha}{\bf{\sigma}}_{\alpha\beta}c_{{\bf
i}\beta}$ is the spin of the mobile electron, the  Pauli
matrices are denoted by ${\bf{\sigma}}$,
${\bf{S}_i}$ is the localized
spin at site ${\bf i}$,
${ \langle {\bf ij} \rangle }$ denotes nearest-neighbor (NN)
lattice sites,
${\rm t}$ is the NN-hopping amplitude for the electrons,
${\rm J>0}$ is an antiferromagnetic (AF) coupling between the spins of
the mobile and localized degrees of freedom,
and ${\rm J'>0}$ is a direct AF coupling
between the localized spins.
The density $\rm \langle n \rangle$=$\rm 1-x$ of 
itinerant electrons is controlled by a chemical potential $\mu$. 
Hereafter ${\rm t=1}$ will be used as the unit of energy. 
${\rm J'}$ and ${\rm J}$ are fixed to 0.05 and 2.0 respectively, 
values shown to be realistic in previous investigations \cite{adri1}. 
The temperature will be fixed to a low value:
T=0.01, which was shown before to lead 
to the correct high-$T_c$ phenomenology \cite{adri1}. 

To simplify the numerical calculations, avoiding the sign problem, the
localized spins are assumed to be classical (with $\rm |S_{\bf i}|$=1).
This approximation is not drastic, and it was already discussed in 
detail in Ref. \cite{adri1}. 
The model will be analyzed using a
Monte Carlo (MC) method, 
details of which can be found in Ref.~\cite{yuno}.
To study the superconducting properties of the system, the pair
correlation functions, 
$C_w({\bf r})=\langle \hat\Delta ^w_{\bf i+r}\hat\Delta ^{w
\dagger}_{\bf i}\rangle$,
are measured. The index $w$ indicates D- or extended S-wave 
pairing, and the
pairing operator is given by
$\hat\Delta ^w_{\bf i}=
\sum_{\alpha}\sum_{\hat e=\pm\hat x,\pm\hat y}\alpha e c_{\bf
i,\alpha}c_{\bf i+\hat e,-\alpha}$,
where $e=1$ for $\hat e=\pm\hat x$ and $ \hat e=\pm\hat y$ for
S-wave pairing, while $e=1$ for $\hat e=\pm\hat x$ and $e=-1$ for
$\hat e=\pm\hat y$ in the case of $d_{x^2-y^2}$-wave pairing.

To study the long distance behavior of the pairing correlations, 
results on $N \times N$ ($N=8$ and 12), 
clusters are here presented. 
These results show that the D-wave correlations
are stronger than S-wave for all the values of the parameters
studied. A typical
comparison between the two correlations is shown in
Fig.1a for $\langle n \rangle$=$0.75$ on a $12 \times 12$ cluster \cite{note1}.
The extended S-wave exhibits strong fluctuations, while the D-wave
results are more robust and  smoother. This shows that our SF-model
captures the essence of hole pairing in AF backgrounds, where it is well
known that $d_{x^2-y^2}$ correlations dominate \cite{Kar}. 
Thus, from now on, we will 
concentrate only on the
behavior of the D-wave pairing correlations \nolinebreak (DPC).

\begin{figure}[thbp]
\centerline{\psfig{figure=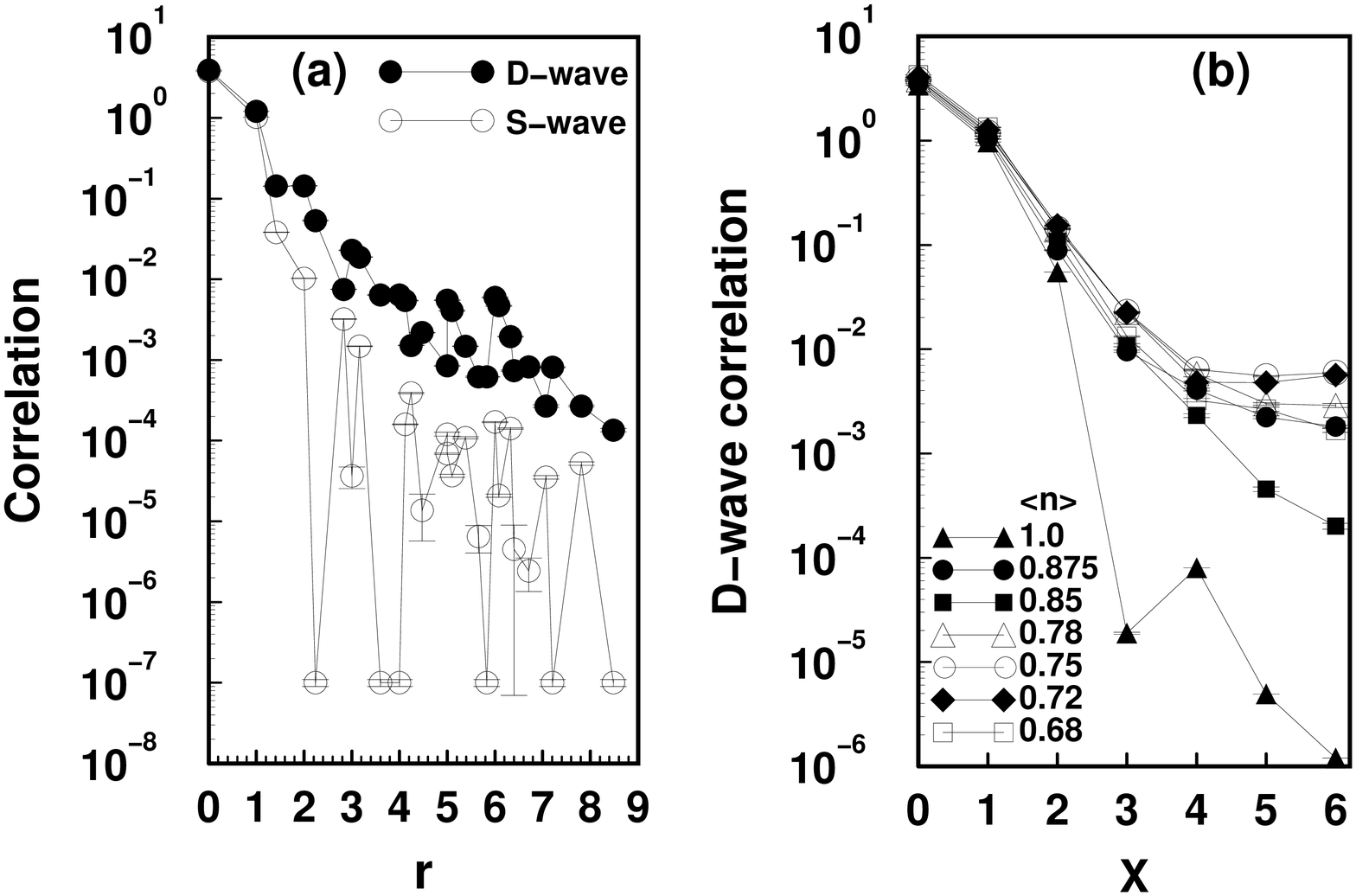,width=8cm}}
\vskip0.5cm
\centerline{\psfig{figure=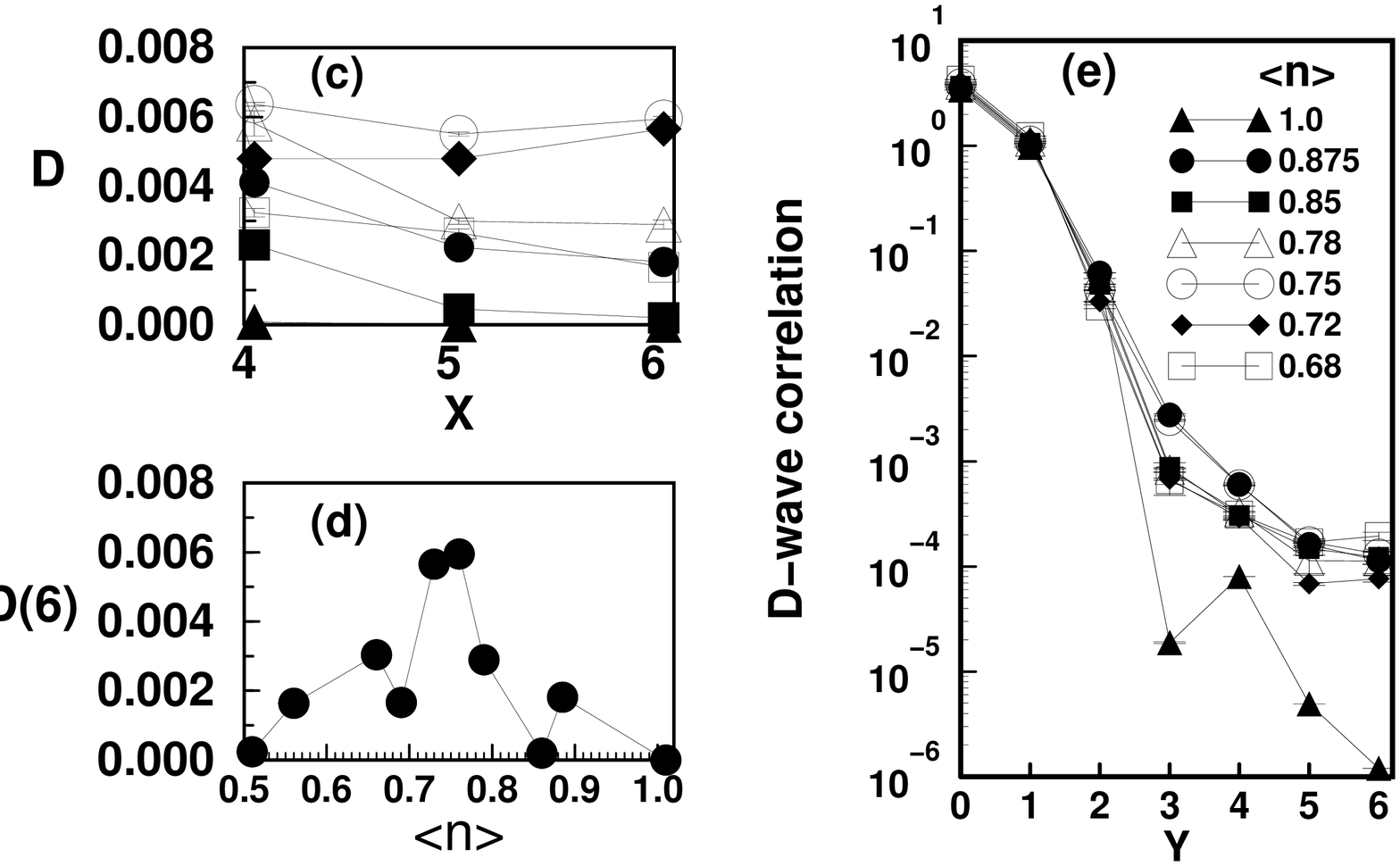,width=8cm}}
\vskip 0.3cm
\caption{(a) D-wave  (filled circles) and extended
S-wave (open circles) pairing correlation versus
distance $r$, on a $12 \times 12$ lattice at 
$\langle n \rangle$=0.75. (b) D-wave pairing correlations
along the direction perpendicular to the charge inhomogeneities at
the densities indicated. (c) The long distance
behavior of the correlations shown in (b), displayed in greater detail.
(d) Correlations for $x$=6, as a function of density.
(e) Same as
(b) but with correlations measured parallel to the
stripe-like charge inhomogeneities. }
\end{figure}

In Fig.1b we present the DPC versus
distance (measured in the direction {\it perpendicular} to the
stripe-like inhomogeneities) for several electronic densities. 
At half-filling the correlations are very small , i.e. $\sim 10^{-6}$,  
at the largest distance, but they develop
a fairly robust tail of order $10^{-2}$ at
small hole
doping (corresponding to an order
parameter $\langle \hat\Delta ^w_{\bf i} \rangle$$\sim$0.1). 
At these densities expecting stronger pair correlations would be
unrealistic, since the low carrier density 
as well as the previously extensively documented small quasiparticle weight Z 
of holes in antiferromagnets
suppresses the signal \cite{bob}. In addition,
the pairing operator, being nearest-neighbors, 
is not optimized to fit the actual pair size.
Ours is among the strongest signals for D-wave SC found 
in unbiased studies of
realistic high $T_c$ models, and they are even comparable to
those reported in 2-leg ladders \cite{elbio}. 
The strongest correlations are observed at $\langle n
\rangle\approx 0.75$, indicating the existence of an optimal doping as in
real cuprates \cite{birg}. This
behavior can be observed in more detail in Fig.1c.
In Fig.1d the DPC at distance $x$=6, is shown as a function of
electronic density, and the existence of an optimal doping is again clear.
The dip at $\langle n
\rangle \sim 0.85$, which 
corresponds to a state that has nearly static stripes, 
can be qualitatively identified to the $x$=1/8
anomaly observed in the cuprates.\cite{tran2}
The correlations in the direction parallel to the stripes are in
Fig.1e and, surprisingly, they are about one order of
magnitude smaller than those in Fig.1b. This is a remarkable general 
trend, and the reasons for this difference are discussed below.

Previous studies have shown that the
system changes from an AF insulator to a metal upon 
doping \cite{adri1}.
The large suppression of the pairing correlations at 
half-filling 
is expected due to the absence 
of holes. However, as observed in Fig.1b, an increase of about 
four orders of magnitude occurs when the system is doped 
and becomes metallic. To understand this phenomenon, 
we will analize the properties of the charge and spin
configurations (MC ``snapshots'')
that contribute the most to the enhancement of the DPC, starting at
densities for which fairly 
static stripes are stabilized.

\begin{figure}[thbp]
\centerline{\psfig{figure=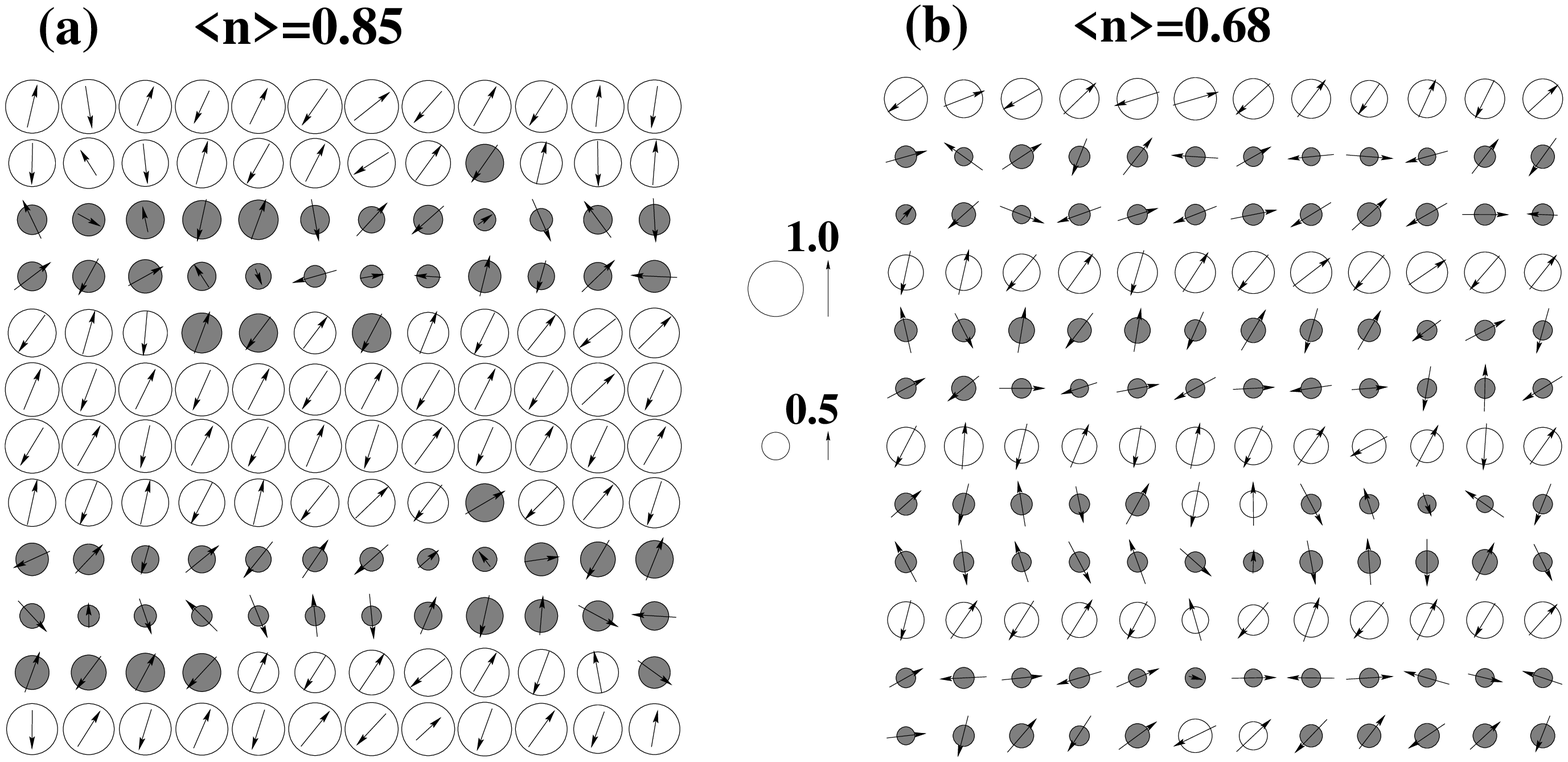,width=8cm}}
\vskip 0.5cm
\centerline{\psfig{figure=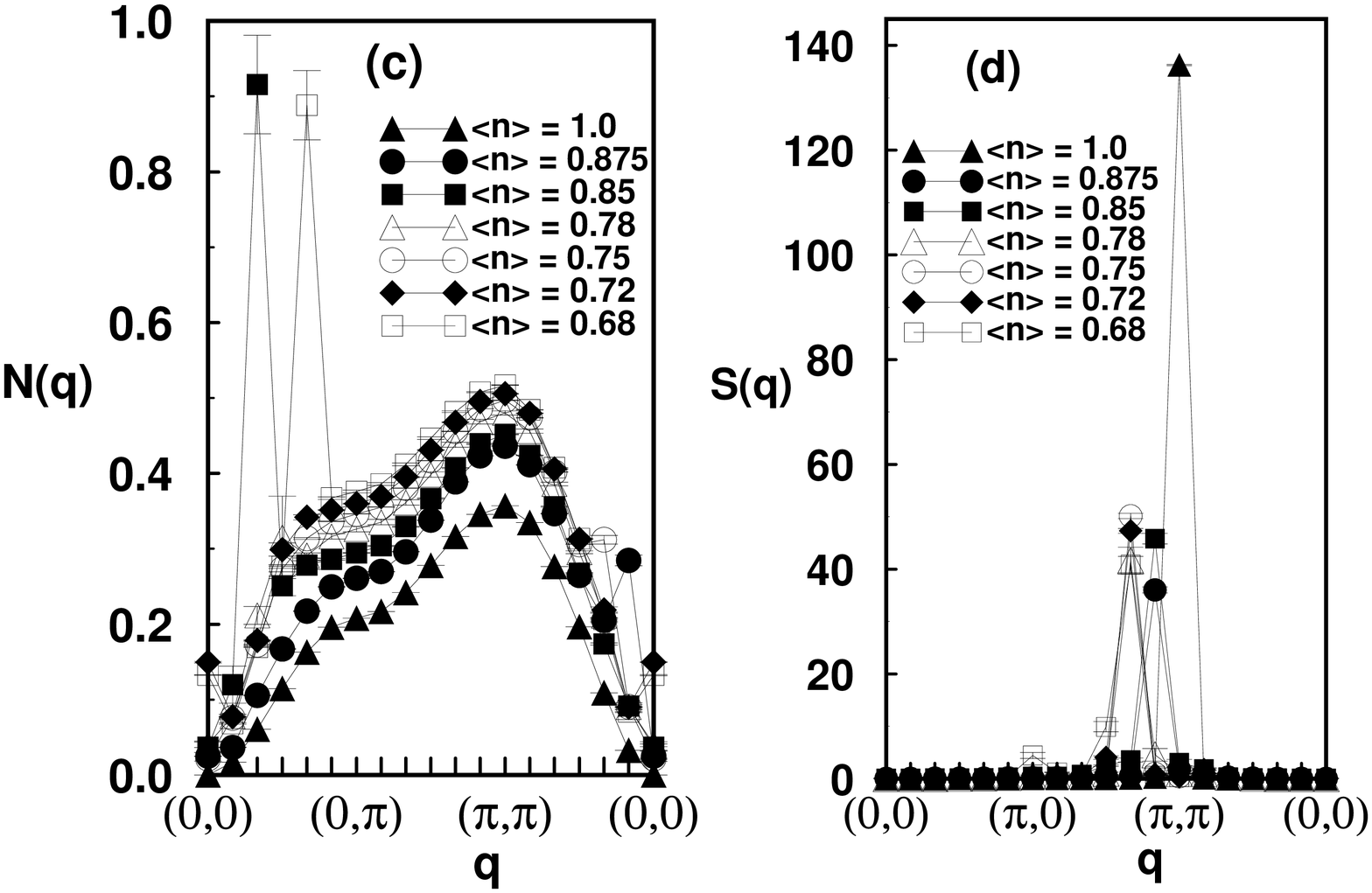,width=8cm}}
\vskip 0.3cm
\caption{(a) Representative snapshot 
of the spin and charge degrees of freedom on a $12
\times 12$ cluster at $\langle n \rangle$=0.85, in the regime of
nearly {\it static} stripes. 
The
area of the circles is proportional to the electronic charge, while the
length of the arrows is proportional to the projection of the localized
spins on the X-Y plane. When $n(i)<\langle n\rangle$ the circles are
gray.
(b) Same as (a) but for $\langle n \rangle$=0.68.
(c) Charge structure factor along selected
directions in momentum space for the same parameters as in Fig.1b. Stripe-like
inhomogeneities for all dopings are along the x direction.
(d) Magnetic structure factor along selected
directions in momentum space for the same parameters as in part (c).}
\end{figure}

The spin and
charge distribution for a typical MC 
snapshot at $\langle n \rangle$=0.85 is shown 
in Fig.2a. When the local density is smaller than the average density
the circles proportional to the local charge density are shown in
gray. As observed, the gray circles determine
two fairly static horizontal stripes \cite{note2}. 
This inhomogeneous charge distribution produces a
large peak at momentum $(0,\pi/3)$ in the charge structure factor $N(q)$, shown
in Fig.2c. It
is also clear the nearly perfect AF order in the electron
rich regions (white circles). The magnetic structure factor $S(q)$, shown in
Fig.2d,  peaks at
momentum $(\pi,5 \pi/6)$ in agreement with the AF order observed
horizontally and the incommensurability induced vertically by the
stripes, which carry a $\pi$-shift across them. 

The results in Figs.1b,e are averages over MC
time, and also over all the sites of the lattice. Measurements 
done directly on the individual snapshots were found to be 
similar to the averages,
and indicate that the pair correlations
are weaken along the AF regions. An intermediate value is obtained along
the hole-rich stripes, but the largest correlation is observed along the
direction perpendicular to the stripes (see Fig.3a).

\begin{figure}[thbp]
\centerline{\psfig{figure=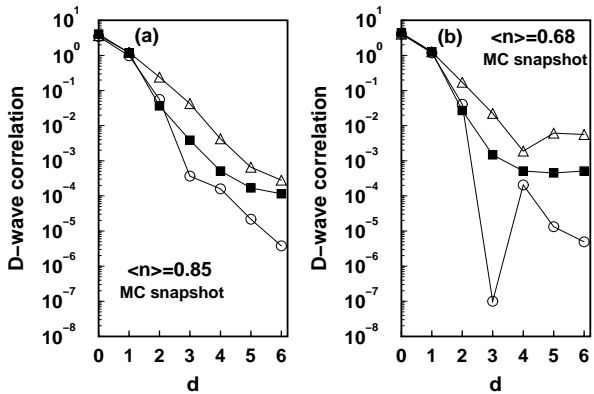,width=8cm}}
\vskip 0.3cm
\caption{(a) D-wave correlations for the representative
snapshot shown in Fig.2a.
The circles indicate correlations along row 6 (counting from the bottom
of Fig.2a), which corresponds to an AF array with local charge $\sim
0.97$. The filled squares are correlations along row 3, which 
has an average local charge $\sim0.67$. The triangles are
correlations in the direction perpendicular to the stripes starting from
row 2 and averaged over all the columns. 
(b) D-wave correlations for the 
snapshot shown in Fig.2b.
The circles indicate correlations along row 3 (counting from the bottom
of Fig.2b), which corresponds to an AF array with local charge $\sim
0.78$. The filled squares are correlations along row 8, which 
has a local charge $\sim0.63$, and the triangles are
correlations in the direction perpendicular to the stripes starting from
row 6. }
\end{figure}

An analogous effect, but more enhanced, is observed 
for $\langle n \rangle$=0.68, Fig.2b, where four nearly static horizontal
stripes are stabilized. 
$N(q)$ (Fig.2c) peaks at momentum $(0,2 \pi/3)$, while $S(q)$
(Fig.2d) at $(\pi, 2 \pi/3)$. In this case,
according to Fig.1b and 1e, the DPC functions are one
order of magnitude larger
in the direction perpendicular to the stripes than in the
parallel direction. Note also that there is a substantial
difference in the DPC perpendicular to the stripes corresponding to the densities 
0.68 and 0.85 here analyzed. To shed some light on these issues,
in Fig.3b correlations for the snapshot shown in Fig.2b are presented.
The correlations (circles) along the AF domains 
parallel to the
stripes, with an average local density of $\sim 0.78$, 
are very weak. We
believe that the AF order
is responsable for this depletion. The DPC
are stronger along the hole-richer stripes, as indicated by the
squares in Fig.3b. In this case, the local density is $\sim
0.63$. However, the strongest correlations, indicated by triangles,
occur, once again, along the direction perpendicular to the stripes. Along this
direction the local charge is very inhomogeneous and magnetic
incommensurability occurs.
These observations lead us to believe that local charge homogeneity and AF
order do not favor D-wave pairing, while local charge inhomogeneity and its
associated magnetic incommensurability promote it.
Thus, the difference in the perpendicular correlations for $\langle n
\rangle$=0.85 and 0.68 mentioned above may be related to the AF reduction 
in electron rich regions, as the system is doped away from half-filling.

\begin{figure}[thbp]
\centerline{\psfig{figure=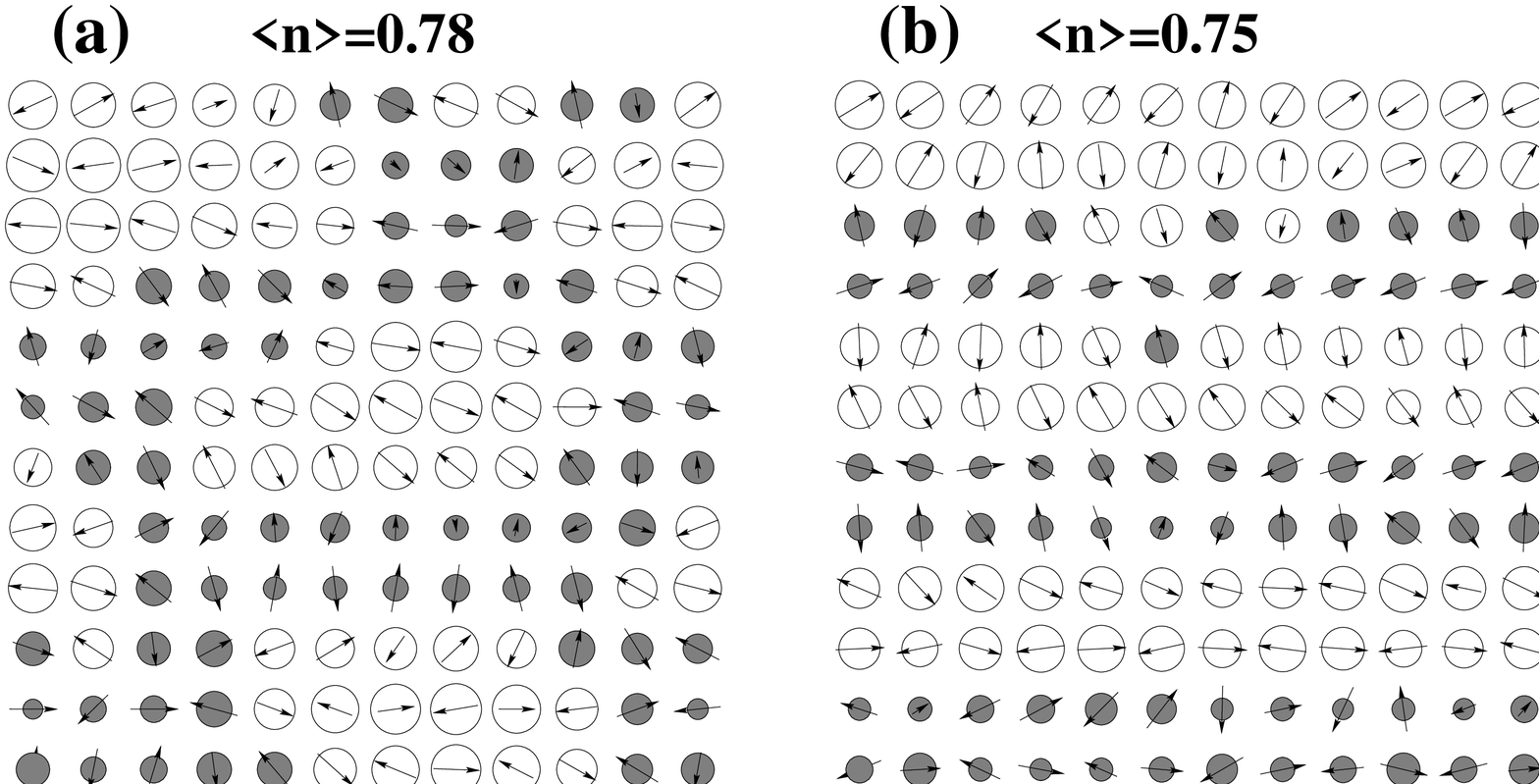,width=8cm}}
\vskip 0.3cm
\centerline{\psfig{figure=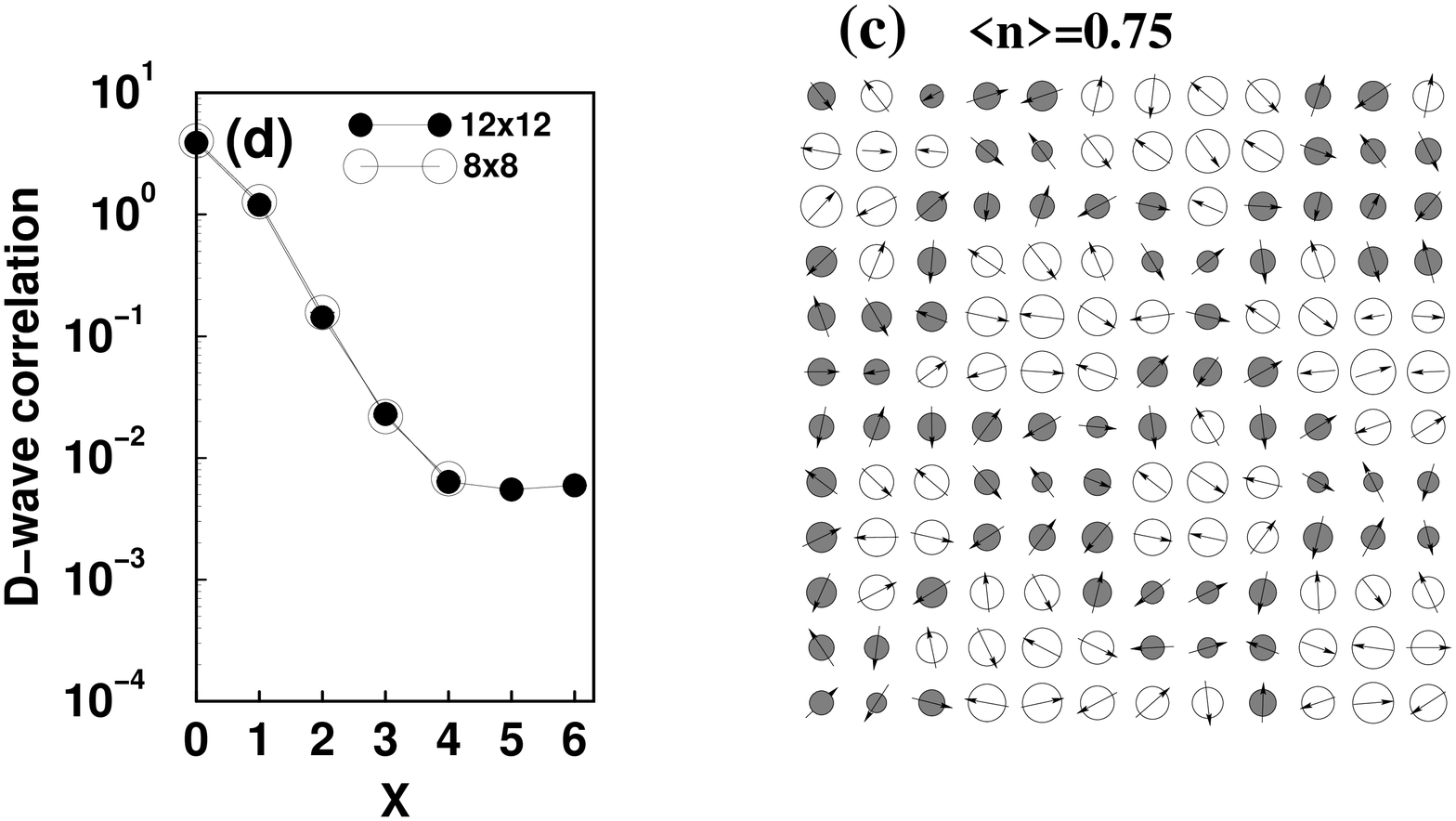,width=8cm}}
\vskip 0.3cm
\caption{(a) Representative snapshot of the spin and charge degrees of freedom on a $12
\times 12$ cluster at $\langle n \rangle$=0.78, in the regime of {\it dynamic} stripes. 
(b) Same as (a) for $\langle n \rangle$=0.75. (c) Snapshot for
the same parameters as in (b) obtained at a later time during the
simulation. (d) DPC
in the direction perpendicular to the stripes using two lattice sizes.}
\end{figure}

The largest DPC  
are obtained for densities where
charge fluctuations
appear to be $dynamical$, according to the snapshots observed,
and their time evolutions,
during several MC runs. This is in agreement with previous expectations
based on experimental results for the high $T_c$ cuprates \cite{tran}.
To further confirm this picture, in Fig.4a a typical 
MC snapshot for $\langle n\rangle$=0.78 is shown. Although 
the charge distribution is clearly inhomogeneous, 
the electron-rich AF domains are not separated by static stripes. 
We have observed that the shape
of these domains changes substantially during the MC
simulation, confirming that the charge structures are dynamical. As
a result, only a small peak at $(0,\pi/2)$ is observed in
$N(q)$ (Fig.2c).
The magnetic incommensurability, on the other hand, exhibits 
a very sharp peak at $(\pi,2\pi/3)$ (Fig.2d). As
observed in
Fig.1b, the pairing correlations are enhanced in the direction
parallel to the spin incommensurability, which is mainly vertical
based on $S(q)$. 
Also in Fig.1b it can be seen 
that the pairing correlations are the strongest for $\langle n
\rangle$=0.75. The corresponding snapshots are in Figs.4b and c. 
As in the previous case, a dynamical inhomogeneous charge distribution is 
observed. Rigid stripes appear at times during the simulation (Fig.4b), but
they become distorted after a few subsequent MC iterations (Fig.4c).
A very weak peak at $(\pi/3,\pi/3)$ exists in $N(q)$
(Fig.2c), while a sharp peak at $(\pi,2\pi/3)$ (Fig.2d) 
is found in its magnetic counterpart.
Similar behavior was observed at $\langle n \rangle$=0.72 
and 0.875.

Our conclusion is that weak charge inhomogeneity and 
magnetic incommensurability
appear to be crucial for the development of robust DPC, at least
within the SF-model.
Although $N(q)$ seems nearly featureless
in the regime with the strongest pairing, the states are $not$ homogeneous,
as observed from the MC snapshots.
In fact studies performed by us in
the non-interacting system show that when the charge is uniformly distributed,
the magnetic structure factor is featureless,
and the DPC present strong fluctuations and are
suppressed \cite{dougf}.

A comparison of results on $8 \times 8$ and
$12 \times 12$ clusters, 
indicates that charge inhomogeneities become more dynamical
as the system size increases. For example, at $\langle n
\rangle$=0.75 the stripes appear static on $8 \times 8$ clusters but, as
shown, are more dynamical on $12 \times 12$ ones.
In Fig.4d, we present the DPC perpendicular to the charge
inhomogeneities for $\langle n \rangle$=0.75 on an $8 \times 8$ and a
$12 \times 12$ cluster. The figure shows that, on one hand, finite size 
effects are small but, on the other hand, a long range tail
starts developing on the $12 \times 12$ cluster and it is not apparent
in the $8 \times 8$ one; its origin appears to be related to the
development of fluctuations in the charge distribution that, as said
above, are observed only in the larger cluster.

Summarizing, we have found evidence of robust D-wave pairing
correlations in a doped SF-model, with a charge inhomogeneous
ground state. Our results indicate that static 
AF order and D-wave pairing
compete. The latter is enhanced when AF is replaced by magnetic
incommensurability. Static, stripe-like charge inhomogeneities, 
only decrease the strength of the pairing correlations, as
compared with the effect of
dynamic charge inhomogeneities. This is reasonable since a ``liquid''
charge distribution should be more favorable to pairing than a
``crystal'' one. In addition, the hole attraction strength caused
by antiferromagnetism should be maximized near half-filling, where 
two holes form a bound state in, e.g., the t-J model \cite{Kar}.
As a consequence, an optimal doping emerges where AF pairing is still robust,
while static stripes do not compete with superconductivity. 
A similar behavior was
observed in Ref.\cite{field}, which provide
evidence that SC can coexist with static stripe order in 
${\rm La_{1.6-x}Nd_{0.4}Sr_xCuO_4}$. However, when this coexistence is
observed, $T_c$ is at a minimum indicating that static stripe order competes
with SC. According to our results, D-wave
SC is expected to be maximized when the charge
inhomogeneities are the most dynamic, a situation that appears 
to occur in doped LSCO. In this situation,
magnetic incommensurability manifest clearly as a peak in the magnetic
structure factor, while $N(q)$ is almost
featureless. This is similar to the behavior observed with neutron
scattering in high $T_c$ cuprates \cite{tran}.

A.M. is supported by NSF under grant DMR-9814350.
Additional support is provided by the National High Magnetic Field Lab 
and MARTECH.


\begin{references}
\bibitem{note1} For the correlations $C_w({\bf r})$ shown, 
${\bf r}=(x,y)$ is such
that $y \geq x$ and the distance $r=\sqrt{x^2+y^2}$.
Qualitatively similar results are found for all possible
distances between the pair correlations in the clusters.
\bibitem{dougf} Our observations in the SF-model
differ from studies of the t-t$'$-J model on ladders where DPC
were enhanced for positive values of the diagonal
hopping t$'$ as the charge order was suppressed, not due to fluctuations
of the domain walls, as in this study, but due to the complete
evaporation of the stripes which made the $\pi$-shifted AF
regions disappear \cite{steve}. It is also found in
Ref.\cite{steve} that when DPC are dominant, the magnetic
correlations peak at $(\pi,\pi)$, in contrast with our results and the
experimental data for the high $T_c$ cuprates in which
magnetic incommensurability is observed when this happens.
\end{references}
\end{document}